\documentclass[a4paper]{jpconf}
\usepackage{graphicx}
\begin{document}
\title{$w$-singularities in cosmological models}

\author{Leonardo Fern\'andez-Jambrina}

\address{ETSI Navales, Universidad Polit\'ecnica de Madrid, Arco de 
la Victoria s/n, E-28040-Madrid, Spain}

\ead{leonardo.fernandez@upm.es}

\begin{abstract}
Recently a new type of cosmological singularity has been postulated
for infinite barotropic index $w$ in the equation of state $p=w \rho$
of the cosmological fluid, but vanishing pressure and density at the
singular event.  Apparently the barotropic index $w$ would be the only
physical quantity to blow up at the singularity.  In this talk we
would like to discuss the strength of such singularities and compare
them with other types.  We show that they are weak singularities.
\end{abstract}

\section{Introduction}
Until the observation of accelerated expansion of our universe, the
only types of singularities that were considered to appear in
Friedman-Lema\^itre-Robinson-Walker (FLRW) cosmological models were
singularities in the form of a big bang or a big crunch, for which the
density of the matter content of the spacetime blows up \cite{HE}, as well as
the scale factor of the universe.
However, since the observational value of the barotropic index of the 
universe nowadays (the quotient between the pressure and the density of the 
content of the universe, $w=p/\rho$) is around minus one, there is a 
possibility of violation of the energy conditions and new types of 
singularities could appear. In fact, such singularities appear 
naturally in gravitational theories which try to cope with the 
observed phenomena.

Many of these new singularities are included in a descriptive
classification due to Nojiri, Odintsov and Tsujikawa (N.O.T. in the
following), which takes into account which physical quantities (scale
factor of the universe $a$, Hubble ratio $H$, pressure $p$ or density
$\rho$) are singular \cite{Nojiri:2005sx}:

\begin{itemize}    
   \item Big bang / crunch: Vanishing $a$, divergent $H$, $\rho$ and 
   $p$.
   
   \item Type I: ``Big rip'': Infinite $a$, $\rho$ and $p$ \cite{Caldwell:2003vq}.

   \item Type II: ``Sudden'': Finite $a$, $H$ and $\rho$, divergent 
   $\dot H$ and $p$ \cite{sudden}.

   \item Type III: ``Big freeze'': finite $a$, infinite $H$, 
   $\rho$ and $p$ \cite{freeze}.

   \item Type IV: Finite $a$, $H$, $\dot H$, 
   $\rho$ and $p$, but infinite higher derivatives of $a$.
\end{itemize}

The strength of these singularities \cite{tipler,krolak} has been 
checked in \cite{puiseux,modigravi}. Summarizing, big rip 
singularities are strong whereas sudden \cite{suddenferlaz}, big freeze and type IV 
singularities are weak. In this sense, the latter cannot be taken as 
the final stage of the universe, since the spacetime could be 
extended continuously beyond the singularity.

Out of this scheme, directional singularities for which some observers
experience an infinite curvature, though the curvature scalars vanish
at the singularity, have been found in some theories \cite{hidden}.

Also out of this classification, in \cite{wsing} a FLRW cosmology is 
shown in which only the barotropic index $w$ becomes infinite at $t=t_{s}$,
\begin{eqnarray}
\label{wmodel}
a(t)&=&\frac{a_s}{1-\frac{3\gamma}{2}\left(\frac{n-1}{n-\frac{2}{3\gamma}}\right)^{n-1}}\nonumber
+\frac{1-\frac{2}{3\gamma}}{n-\frac{2}{3\gamma}}\frac{na_s}{1-\frac{2}{3\gamma}\left(\frac{n-\frac{2}{3\gamma}}{n-1}\right)^{n-1}}\left(\frac{t}{t_s}\right)^{\frac{2}{3\gamma}} \nonumber
\\&+& \frac{a_s}{\frac{3\gamma}{2}\left(\frac{n-1}{n-\frac{2}{3\gamma}}\right)^{n-1}-1}\left(1-\frac{1-\frac{2}{3\gamma}}{n-\frac{2}{3\gamma}}\frac{t}{t_s}\right)^n~,
\end{eqnarray}
for $\gamma>0$ and $n\neq1$ a natural number, whereas the scale factor, $a(t_{s})=a_{s}$,
and all its derivatives remain finite and the density and the 
pressure vanish at $t_{s}$. In \cite{yurov} similar singularities are obtained but with 
non-vanishing pressure.

This new type of singularities has been dubbed barotropic index 
$w$-singularities \cite{wsing}, or simply $w$-singularities, or even 
type V \cite{kiefer} singularities, remarking the difference between 
them and type IV singularities, since no higher derivatives of the 
scale factor blow up at the time of the singularity $t_{s}$.

In this talk we would like to show a characterization of barotropic 
index $w$-singularities in FLRW cosmologies in terms of the form of 
the scale factor. More details may be found in \cite{wchar}.

In the next section we derive a characterization of barotropic index
$w$-singularities in terms of the exponents of the power expansion in
time of the scale factor of the universe.  A discussion of the results
is included in the last section.

\section{Characterization of barotropic index $w$-singularities}
In gravitational theories such as general relativity, the content of the universe is depicted as 
a perfect fluid of density $\rho$ and pressure $p$, depending on just 
the time coordinate $t$, due to the homogeneity and isotropy of FLRW 
spacetimes. This means that there is an equation of state for the 
perfect fluid, $p=p(\rho)$, and the ratio of pressure and density is 
defined as the barotropic index of the fluid, $w=p/\rho$. This index 
evolves with time except for power-law cosmologies, which have a 
constant barotropic index.

In flat FLRW cosmologies the barotropic index is written in terms of 
the scale factor of the universe and its derivatives,
\begin{equation}\label{prho}
\rho=3\left(\frac{\dot a}{a}\right)^2,\qquad
p=-\left(\frac{\dot a}{a}\right)^2-\frac{2\ddot
a}{a}.\end{equation} 
\begin{equation}
w=-\frac{1}{3}-\frac{2}{3}\frac{a\ddot a}{\dot a^2},\end{equation}
and if we assume that it may be expanded as a generalized power 
expansion of time around the time of the singularity \cite{visser,puiseux} 
\begin{eqnarray}\label{expan}
a(t)=c_{0}(t_{ s}-t)^{\eta_{0}}+c_{1}(t_{ 
s}-t)^{\eta_{1}}+\cdots,\qquad
\eta_{0}<\eta_{1}<\cdots,\qquad c_{0}>0,
\end{eqnarray}
where $\eta_{0}$, $\eta_{1}$,\ldots are real numbers, 
the barotropic index can be expanded accordingly,
\[w(t)=w_{0}(t_{ s}-t)^{\zeta_{0}}+w_{1}(t_{s}-t)^{\zeta_{1}}+\cdots,\]
and the exponents and coefficients of this expansion can be written in terms 
of the former ones.

This gives rise to a hierarchy of cases and subcases depending on the 
values of $\eta_{0}$ and $\eta_{1}$ \cite{wchar} that are consigned 
in table~\ref{tablw}. We notice that infinite barotropic indices 
appear just for vanishing $\eta_{0}$ exponent, except for 
$\eta_{1}=1$ with $\eta_2\ge 2$.
\begin{table}
\caption{Singularities in  barotropic index.\label{tablw}}
\begin{center}
\begin{tabular}{llll}
   \br
   ${\eta_{0}}$ & ${\eta_{1}}$ & $\eta_{2}$ &$w_{ s}$ \\
   \mr
   Nonzero & $(\eta_{0}, \infty)$ &   $(\eta_{1}, \infty)$ & Finite \\ 
   Zero & $(0,1)$ &   $(\eta_{1}, \infty)$ &   Infinite  \\
      & $1$ &  $(1,2)$ & Infinite  \\
      & $1$ & $[2, \infty)$ & Finite  \\
      & $(1, \infty)$ & $[\eta_{1}, \infty)$ & Infinite \\
   \br
   \end{tabular}\end{center}
\end{table}

The requirement of finite derivatives of the scale factor implies 
that all exponents $\eta_{i}$ must be natural numbers. Hence, we are 
left just with the subcase of infinite barotropic index with 
$\eta_{0}=0$ and $\eta_{1}\ge 2$.

From (\ref{prho}) and (\ref{expan}) we learn that
vanishing density requires $\eta_{1}>1$ and vanishing pressure
requires $\eta_{1}>2$ \cite{wchar}.

Hence we see that an infinite barotropic index with finite 
derivatives of the scale factor of the universe requires a vanishing 
density at the time of the singularity but not a vanishing pressure.

This may be summarized in the following statement:\\

\noindent\textbf{Theorem:} A FLRW cosmological model has an infinite
barotropic index at a finite time $t_{ s}$ with finite derivatives of
the scale factor of the universe if and only if $a(t)$ can be 
expanded as a Taylor series at $t_{s}$ with a null linear term,
\begin{equation}a(t)=c_{0}+\sum_{n=2}^{\infty}c_{n}(t_{s}-t)^{n}.\end{equation} 
Furthermore if we require a vanishing pressure either (barotropic 
index $w$-singularity), the quadratic term $c_{2}$ is also to be zero.
\section{Conclusions \label{discuss}}

We have shown a simple characterization of barotropic index 
$w$-singularities in FLRW cosmological models in terms of Taylor 
expansions of the scale factor of the universe. The necessary and 
sufficient condition is just the vanishing of the linear term of the 
expansion. And if we consider singularities with vanishing pressure, 
then both the linear and quadratic terms must be zero.

Most exotic singularities are weak in the sense that when we 
consider finite objects they are not disrupted by tidal forces on 
experiencing the singularity, since the spacetime can be extended 
continuously beyond the singular event \cite{puiseux}.

In that reference, the strength of the singularities was checked in
terms of the exponents of the expansion (\ref{expan}) of the scale
factor of the universe.  Though barotropic index $w$-singularities
were not considered, the results obtained there are general enough to
be applied to cosmological models with vanishing linear term.  In
fact, it was shown that singularities in cosmological models with
$\eta_{0}=0$ and $\eta_{1}$ larger or equal than one were weak
\cite{clarke} with both Tipler \cite{tipler} and Kr\'olak's
\cite{krolak} criteria.

Hence barotropic index $w$-singularities are weak and cannot be 
considered as a final stage for the universe, as it happens with 
other exotic singularities but big rip. 

In table~\ref{scale} we summarize our results on exotic singularities
and their strength in terms of the exponents of the power expansion of
the scale factor of the universe around the time of the singularity.

Singularities in the derivatives of the scale factor of the universe
appear only for non-natural exponents in the expansion.  This is the
reason for including a column $\{\eta_{i}\}$ which deals with that
fact.  

The letter I (independent) means no additional condition on the 
exponents of the expansion around the time of the singularity. S (some)
states that there must be at least one non-natural exponent in order
to have a singularity in one of the derivatives. Finally, N (natural) means that all exponents are natural.

\begin{table}
\caption{Strength of singularities in  FLRW cosmological models\label{scale}.}
\begin{center}
\begin{tabular}{llllllll}
   \br
   ${\eta_{0}}$ & ${\eta_{1}}$ & $\eta_{2}$ & $\{\eta_{i}\}$ &\textbf{Tipler} &
   \textbf{Kr\'olak} & \textbf{N.O.T.} \\\mr
   $(- \infty,0)$ & $(\eta_{0}, \infty)$ &   $(\eta_{1}, \infty)$ & I&
   Strong & Strong  & I\\ 
   $0$ & $(0,1)$ &   $(\eta_{1}, \infty)$ & S&  Weak & Strong & III \\
      & $1$ & $(1,2)$ & S& Weak & Weak & II \\
      &  & $[2, \infty)$ & S & Weak & Weak & IV \\
      & $(1,2)$ &  $(\eta_{1}, \infty)$ & S & Weak & Weak & II \\
      & $[2, \infty)$ &   $(\eta_{1}, \infty)$ & S&  Weak & Weak
      &  IV \\      & 2 &   $[3, \infty)$ & N & Weak & Weak
      &  $w$ with $p_{s}\neq0$\\      & $[3, \infty)$ &   $[\eta_{1}+1, \infty)$ & N & Weak & Weak
      &  $w$ with $p_{s}=0$\\ $(0, \infty)$ & $(\eta_{0}, \infty)$ &   $(\eta_{1}, 
      \infty)$ & I&
Strong & Strong & Crunch\\
   \br
   \end{tabular}
\end{center}
\end{table}

\section*{Acknowledgments}The author wishes to thank the University of
the Basque Country for their hospitality and facilities to carry out
this work.

\end{document}